\documentstyle[11pt,aas2pp4]{article}
\input{psfig}
\journalid{}{}
\articleid{}{}
\def\simlt{\mathrel{\hbox{\rlap{\hbox{\lower4pt\hbox{$\sim$}}}\hbox{$<$}}}}
\def\simgt{\mathrel{\hbox{\rlap{\hbox{\lower4pt\hbox{$\sim$}}}\hbox{$>$}}}}

\begin{document}
\title{ELEMENT RATIOS AND THE FORMATION OF THE STELLAR HALO}
\author{Gerard Gilmore}
\affil{Institute of Astronomy, University of
Cambridge, Madingley Road, Cambridge, UK CB3 0HA}
\author{Rosemary F.~G. Wyse}
\affil{Department of Physics and Astronomy, The Johns Hopkins University,
    Baltimore, MD 21218}

\begin{abstract}
It is well-established that the vast majority of metal-poor Galactic
halo stars shows evidence for enrichment by solely massive
stars. Recent observations have identified more varied behavior in the
pattern of elemental abundances measured for metal-rich, [Fe/H]$\sim
-1$~dex, halo stars, with both high and low values of the ratio of
[$\alpha$/Fe].  The low values are most naturally due to the
incorporation of iron from Type Ia supernovae.  These `low-alpha' halo
stars have been interpreted as being accreted from dwarf
galaxies. However, these stars are on very high-energy radial orbits,
that plunge into the Galactic center, from the outer regions of the
stellar halo. We demonstrate here that known dwarf galaxies could not
reproduce the observations, which require the proposed parent dwarf to
have a very high mean density to provide `low-alpha' stars on orbits
of such low values of periGalactic distance. Rather, the observations
are consistent with fragmentation of a gaseous proto-halo into
transient star-forming regions, some of which are sufficiently dense
to survive close to the Galactic center, and self-enrich to relatively
high metallicities.  Those that probe the outer halo sustain star
formation long enough to incorporate the ejecta from Type Ia
supernovae.

Further, we point out that the values of the `Type II plateau' in the
element ratios for disk stars and for halo stars are equal.  This
implies that the (massive) stars that enriched the early disk and halo
had the same IMF.  However, the bulk of the halo stars with
[Fe/H]$\sim -1$~dex that have been observed have significantly lower
values of [$\alpha$/Fe] than does a typical disk star of that
metallicity.  Thus there is a discontinuity in the chemical enrichment
history between the halo and disk, as probed by these samples,
consistent with previous inferences based on angular momentum
considerations.
\end{abstract} 

\keywords{Galaxy: halo --- Galaxy: formation --- Galaxy : evolution --- 
Galaxy --- abundances}

\section{Introduction}

Elemental abundances contain significantly more information than just
`metallicity', since different elements are synthesised in stars of
different masses, and on different timescales.  In particular, oxygen and
the alpha-elements are produced predominantly by massive stars, ejected into
the interstellar medium by  core-collapse, Type II, supernovae.  Iron has
important contributions by both Type II supernovae and Type Ia supernovae,
the latter believed to result from the explosive nucleosynthesis of
degenerate material, associated with a  white dwarf in a binary system.
The vast majority of local stars belonging to the Galactic halo population
have enhanced values, compared to that of the Sun, of the ratio of the
abundances of oxygen and of the alpha-elements to that of iron (e.g. 
Wheeler, Sneden \&  Truran 1989; Nissen et al. 1994; Gratton et
al. 1997a). Further, these halo stars have very similar values of this
`over-abundance' of alpha-elements, [$\alpha$/Fe]$_{halo} \sim +0.3$. This
`halo' value is, within nucleosynthetic yield uncertainties, just that
predicted for chemical enrichment by a mean, IMF-averaged, Type~II (core
collapse) supernova, assuming a normal (i.e. solar-neighborhood)
massive-star mass function (Wyse \& Gilmore 1992; Nissen et al.  1994).

Thus a typical halo star shows no evidence for nucleosynthetic
products from Type Ia supernovae.  This lack restricts the duration of
star formation that created these stars to be shorter than the time it
takes for there to have occurred significant Type Ia explosions.  This
timescale is rather model-dependent.  In the commonly-discussed
double-degenerate models (Iben \& Tutukov 1984; see Iben et al. 1997
for a recent discussion, including the expected statistics of close WD
binaries) it is plausibly of order 1~Gyr after the formation of the
progenitor stars (e.g.  Smecker-Hane \& Wyse 1992; Matteucci \& Fran\c
cois 1992).  A similar timescale for significant supernovae follows
from models in which the white dwarf is accreting due to Roche-Lobe
overflow from a low-mass, evolved companion (Yungelson \& Livio 1998).
This restriction on the duration of star formation is not necessarily
the same as a restriction on the global age range of the halo, since
different regions of the halo could initiate (short-lived) star
formation at different times, resulting in a significantly larger
total age range in the halo (see also Gilroy et al. 1988 for an
analogous argument for rapid local enrichment, based on the appearance
of $s-$process elements in metal-poor halo stars). It is very
implausible that the duration of star formation across the entire halo
was as short as 1~Gyr, no matter one's preconceptions of how the halo
formed -- if `monolithic', this requires remarkable synchronisation
over $\sim 50$~kpc, while if `chaotic', the regions that formed the
halo had remarkably uniform star formation properties.

 Star formation is likely to occur in transient overdense regions,
perhaps of the mass of present-day Giant Molecular Cloud complexes in
the Galactic disk, even in models that envisage a well-defined
`monolithic' Galactic potential well during the formation of the
stellar halo (e.g. Fall \& Rees 1985). Was there pre-enrichment of one
region by another?  Did some regions self-enrich sufficiently long to
incorporate the products of Type Ia supernovae?  Isochrone-based
determinations of the ages of field halo stars are of limited
accuracy, and indeed the only technically-feasible means of detecting
an extended duration of star formation that may be as short as a Gyr
is to identify halo stars with evidence for pre-enrichment by Type Ia
supernovae.  The manifestation of this is lower values of
[$\alpha$/Fe] than can be understood from Type~II supernovae alone.
Given the expectation of some increase in metallicity with time as
star formation proceeds, the most natural place to look for such an
effect is among the most chemically-enriched halo stars.

We note that a variation in the massive-star IMF can also provide for
lower values of element ratios, still with enrichment by only Type~II
supernovae.  However, there is firm observational evidence for
universality of the IMF, surprising as this may be (see papers in
Gilmore \& Howell 1998), and indeed any invoked IMF variation to
explain lower values of [$\alpha$/Fe] would be extremely {\it ad hoc}.
Thus we favor interpretations within the framework of a fixed IMF.

Several recent analyses, most notably that by Nissen \& Schuster
(1997), have discovered exactly the element ratio behavior expected
for some enrichment by Type Ia supernovae.  The authors of these
analyses have favored, as the explanation for `low-alpha' halo stars,
accretion of stars from distinct objects such as satellite dwarf
galaxies.  Here we argue rather that the observations are more
naturally explained by the variation in the duration of star formation
in regions of the proto-halo. Fragmentation into cool, dense
self-gravitating regions is fairly generic to models of the collapse
of galaxy-scale perturbations; assimilation and mixing of these
transient regions is not to be thought of as accretion.  Of course,
accretion, in the sense of a distinct other proto-galaxy merging into
our proto-galaxy, remains a very attractive explanation for the
creation of the thick disk (e.g. Gilmore \& Wyse 1985; see Velazquez
\& White 1998, for a recent simulation), some $\sim 12$~Gyr ago
(e.g. Gilmore, Wyse \& Jones 1995).

\section {The Observations of Elemental Abundances}

Halo stars with low values of the [$\alpha$/Fe] element ratios were
identified in a study of metal-rich halo stars, those with [Fe/H] $ \sim
-1$~dex, by Nissen \& Schuster (1997), and serendipitously in samples of
local, low-metallicity halo stars by King (1997) and by Carney et al. 
(1997). Nissen \& Schuster studied  approximately equal numbers of `disk'
and `halo' dwarfs (16 disk, 13 halo; so designated based on kinematics) in
the metallicity range $-1.3 \simlt \mathrm [Fe/H] \simlt -0.4$. Two giant
stars in each of the `younger' (e.g. Richer et al.  1996) globular
clusters Ruprecht~106 and Pal~12 were analysed by Brown, Wallerstein  \&
Zucker (1997) and found to have approximately solar values of
alpha-element-to-iron ratios.\footnote{Note that the relative ages of
globular clusters are generally derived with fixed -- enhanced -- values of
the alpha-element ratios, so this result, if a property of all stars in the
cluster, increases the age assigned to these clusters and thus reduces the
age spread within the system of Galactic globular clusters.} 

Figure~1 shows the [Mg/Fe] data, as a function of [Fe/H], for the
Nissen \& Schuster (1997) sample of disk and halo stars, plus the
anomalous metal-poor halo stars from King (1997) and Carney et al.
(1997); the shaded region indicates the locus of the majority of
metal-poor halo stars (e.g. Nissen et al.  1994; Gratton et al.
1997a; note that there are sometimes zero-point offsets in the value
of [Mg/Fe] between different observational data sets, which can lead
to apparent gradients in [$\alpha$/Fe] plotted against [Fe/H] if all
published data are naively combined. Each data set shows similar
behavior in their regions of overlap in [Fe/H].  Since it is probable
that these offsets reflect differences in analysis methods rather than
real systematic variations in [$\alpha$/Fe] as a function of [Fe/H],
we do not consider them further here).

As may be seen in the Figure, the majority of the metal-rich halo
stars observed shows significantly lower values of [$\alpha$/Fe] than
the enhanced, super-solar values of the alpha-element ratios derived
for the more metal-poor halo stars.  Thus the halo stars with enhanced
element ratios are `anomalous' in this metallicity range.  This
contrasts with the typical (metal-poor) halo where the stars with low
values of [$\alpha$/Fe] are the anomalous ones. The halo stars of all
metallicity with low values of [$\alpha$/Fe] share one property --
they tend to be on orbits with large values of apoGalactic distances,
$R_{apo} \simgt 10$~kpc, whereas the `normal' halo stars are not to so
biased to the outer halo. This trend suggests a real physical meaning
to the present orbital parameters. Further, all the halo stars of the
Nissen \& Schuster (1997) sample are on derived orbits with very low
periGalactic distances, $R_{peri} \simlt 1$~kpc; this is consistent
with the apparent correlation between $R_{peri}$ and metallicity
established for such kinematically-selected samples by Ryan \& Norris
(1991)\footnote{Carney et al. (1996) argue that this correlation
arises only when the azimuthal velocity is used to isolate halo stars
and interpret it as reflecting the existence of a separate population
of very metal-poor `far' halo stars on retrograde orbits.  While this
is intriguing, the significance of a distinct population on retrograde
orbits is difficult to establish given that the angular momentum sign
of the orbits of fully 38\% of halo stars will be retrograde, given a
mean azimuthal streaming velocity of +30 km/s, and a V-dispersion of
100 km/s, as found by Ryan \& Norris (1991).}, such that the more
metal-rich halo stars tend to be on orbits of smaller periGalactic
distances.  Thus all the metal-rich halo stars observed are on orbits
with small values of $R_{peri}$, while of these, the stars with low
values of [$\alpha$/Fe] are on orbits with large values of $R_{apo}$.

Note that it is {\it not\/} the case that all metal-poor halo stars with
large ${ R_{apo}}$ necessarily have low values of [$\alpha$/Fe], although
such a trend exists for the metal-rich stars with low values of
[$\alpha$/Fe].  As demonstrated  in Table~10 of Carney et al.  (1997) 
there are many metal-poor stars that have orbits associating them with the
outer halo, but that have normal, enhanced, values of the alpha-element
ratios. The metal-poor halo stars with low values of [$\alpha$/Fe] are on
retrograde orbits (King 1997; Carney et al.  1997), but there is no such
signature in the kinematics of the metal-rich stars with low values of
[$\alpha$/Fe].

\section {What Does It All Mean?}

We wish to understand why a small fraction of stars in the stellar halo,
biased to metal-rich stars on high-energy, high-eccentricity orbits, have 
low values of [$\alpha$/Fe]. 

Consider the pattern of element ratios in  self-enriching systems (of fixed
IMF) with given star formation rates. The solid lines in Figure~1 illustrate
schematically the trends produced in simple (homogeneous) models of chemical
evolution with differing star formation rates.  The important feature is the
location of the turndown due to iron from Type Ia supernovae -- since the
{\it shortest\/} pre-explosion lifetime of Type Ia supernovae is presumed to
be a fixed value, the location of this turndown occurs at a higher value of
[Fe/H] for a higher star formation rate, reflecting the more rapid
metallicity increase with time (note that the effect of gas flows on the
location of the turndown from the `Type II plateau' at early times/lower
metallicities can be incorporated by a suitably-modified star-formation
rate; inflow of unenriched material will reduce [Fe/H] keeping [$\alpha$/Fe] 
fixed.  Outflow driven by Type II supernovae will behave similarly). 
Such an effect has been proposed to explain the element ratios seen
in local disk stars, born at different locations in the disk (Edvardsson
et al.  1993). 

\begin{figure}
\psfig{figure=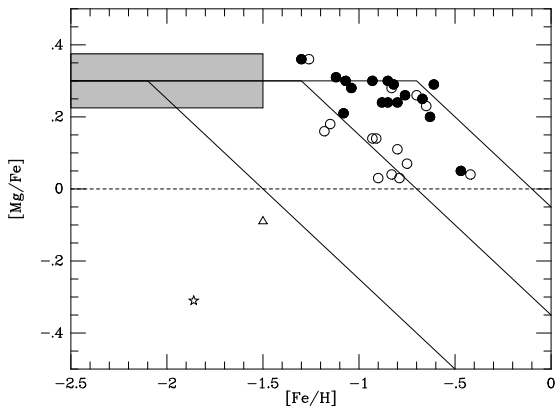,height=6cm} 
\figcaption{ Iron abundance
against element ratio [Mg/Fe] for the halo stars (open circles) and
disk stars (filled circles) from Nissen \& Schuster (1997; their Table
4).  The disk stars show a well-defined trend, contrasted with the
apparent scatter of the halo stars.  The shaded region marks the locus
of normal metal-poor halo stars in this plane. The star symbol
represents the anomalous metal-poor halo subgiant of Carney {\it et
al.} (1997), and the open triangle the anomalous common proper-motion
pair of halo dwarfs from King (1997).  The solid lines schematically
illustrate the expected pattern of element ratios in self-enriching
systems of constant IMF, but varying star formation rates.}

\end{figure}

Figure~1 shows that the observed  differences in element ratio patterns are
inevitable consequences of a variation of the star formation rate.  The
amplitude of the invoked variation in star formation rates is required only
to be such as to allow up to $\simlt 1$~Gyr longer duration of star
formation to produce lower values of [$\alpha$/Fe].   This difference in age
ranges is undetectable with current isochrone-based techniques. Note that
there is no requirement from the element ratio data that the progenitor
stars of the Type Ia supernovae be part of the same star forming region as
the gas they will eventually enrich.   However, the observed trend for low
values of [$\alpha$/Fe] to be observed for higher-metallicity halo stars favors
rather self-enrichment scenarios, in (sub)structure of relatively deep
potential wells. 

Clearly, the data are  consistent with the lower [$\alpha$/Fe] stars forming
over longer timescales, allowing incorporation of iron from Type Ia
supernovae. But where did these halo stars form?  In a distinct separate
galaxy, or essentially where we observe them now? 

\subsection{Accretion Models} 

Nissen \& Schuster (1997) considered two possibilities, either that
the outer regions of the stellar halo of the Galaxy had a longer
duration of star formation than did the inner halo, or that the halo
stars with low values of [$\alpha$/Fe] were accreted from a dwarf
galaxy within which star formation is presumed to have been sustained
over a sufficiently long period, longer than in the field halo.  The
authors favor the latter explanation. Indeed, `distinct accretion
events' are the preferred explanation of all the authors for the data
in Figure~1, and also for the `anomalous' globular clusters (it has
been suggested that these two clusters have been captured from the
Large Magellanic Cloud; Lin \& Richer 1992).

Motivation for the interpretation in terms of `accretion' came
partially from the prediction of low values of the alpha-element to
iron ratios in the stars of dwarf galaxies in which there is extended
star formation, perhaps even successive bursts of star formation, as
inferred for many of the extant satellite galaxies associated with the
Milky Way (Gilmore \& Wyse 1991; Unavane, Wyse \& Gilmore 1996). It is
clear from observations that only the more massive dwarf systems can
self-enrich to a metallicity of 1/10 the solar value (see, for example
the metallicity--luminosity (mass) relation of Lee et al.  1993).  Of
the local dwarf galaxy companions to the Milky Way, only the Fornax
dSph, the Sagittarius dSph and the Magellanic Clouds have a
significant population of stars as metal-rich as the observed
low-alpha halo stars. However, these systems also contain stars of a
wide range of ages (e.g. Smecker-Hane 1996), and late accretion of
stars from systems like these satellite galaxies would produce stars
in the halo which are {\it many\/} Gyr younger than the bulk of the
halo, reflecting the star-formation history of the satellite galaxy
parent. The present sample of `low-alpha' halo stars shows no obvious
signature of having a relative youth of this amplitude, which would be
manifest in their colors (e.g.  Unavane, Wyse \& Gilmore 1996).

Thus late accretion from systems like the extant satellite galaxies is not
favored.  What other kind of accretion is possible? 

Models of the formation of the stellar halo generically invoke early
star formation in transient $\sim 10^7$~M$_\odot$ local potential
wells, perhaps reflecting the initial primordial fluctuation power
spectrum, or formed by thermal and/or gravitational instability within
a larger system.  The turn-around, collapse and relaxation to
dynamical equilibrium of the present-day Galactic potential takes at
least a couple of global dynamical times, or a few Gyr.  Thus
disruption and assimilation of transient systems during the relaxation
of the large-scale Galactic potential should not be considered as
`accretion', since these systems never really had a separate identity.
Systems that do retain coherence through this period can be said to be
`accreted' if assimilated subsequently, and it is these we consider
next.  Since we are discussing essentially old stars, we shall focus
on the stellar components of this substructure; in any case a
satellite galaxy would be most likely stripped of its gas after a few
passages through the gaseous disk of the Milky Way with periGalactic
distances like those observed for the halo stars.

`Substructure' orbiting within the (dark halo) potential of a larger
`parent' galaxy is subject to tidal forces, and to dynamical friction.
The mass and the (mean) density of the substructure will to a large
extent determine its fate.  Dynamical friction will cause the orbit of
the substructure to decay in radius, on a timescale inversely
proportional to the mass of the substructure (for given orbit and
large galaxy); relatively massive substructure will sink to the center
of the larger galaxy on a few orbital times.  Tidal stripping will
cause the substructure to lose material when its internal gravity is
overcome by the gravitational field of the larger galaxy; the simple
Jacobi criterion implies that a system will lose material unless (or
until) its internal density is of order three times the mean density
of the larger galaxy interior to its orbit; simulations have shown
that this simple criterion works remarkably well (e.g. Johnston,
Hernquist \& Bolte 1996).  Diffuse substructure will lose material
while orbiting even in the far outer regions of the larger galaxy.
The orbits of stars that have been accreted into the Galaxy by tidal
effects on a satellite galaxy are only slightly different from that of
the satellite at the time the stars were captured; this produces a
kinematic signature of moving groups of stripped stars, that can
persist for many orbital times (cf. Tremaine 1993; Johnston 1998).

Thus one may consider four broad classes of dissipationless (stars
plus dark matter) substructure: massive, dense systems will survive
fairly intact while sinking to the center of the larger galaxy;
massive, diffuse systems will be largely disrupted by tides during
their orbital decay; low mass, dense systems will be largely
unaffected, while low mass, diffuse systems will be disrupted with no
orbital decay.

The stars under consideration here are on rather eccentric orbits, but
with even $R_{apo}$ much less than that of typical extant satellite
galaxies.  As discussed by Velazquez \& White (1998), for example, in
their study of the interaction between satellite galaxies and large
disk galaxies, {\it local\/} application of the standard
(Chandrasekhar) dynamical friction formula provides reliable results
in the case of non-circular orbits.  Thus due to the higher densities
in the central regions of the bigger galaxy, the frictional effects
are higher there. Further, most tidal stripping occurs at pericenter
passage (e.g. Velazquez \& White 1998 and references therein).  Thus
if the observed halo stars were removed from a satellite galaxy by
tidal stripping, their very low periGalactic distances would require
that the satellite galaxy be able to penetrate deep inside the Milky
Way.  Their low values of $R_{apo}$ also would, in this scenario,
require a rather massive satellite, so that dynamical fraction can
operate on less than a Hubble time.

The periGalactic distances provide the more stringent constraint. As
discussed above, the low [$\alpha$/Fe] stars are on radial orbits,
with periGalactic distances $\simlt 5$~kpc, and indeed the majority of
the stars having $R_{peri} \simlt 1$~kpc.  A lower limit to the mean
density of the Milky Way Galaxy interior to 1~kpc is given by adopting
just the mass of the central bulge, $\sim 1 \times 10^{10}M_\odot$
(e.g. Kent 1992). The predominant stellar populations in the Galactic
bulge and inner disk are probably at least $\sim 8$~Gyr old (Ortolani
et al.  1995; Holtzman et al.  1993), consistent with theoretical
expectations for the timescale of assembly of the `core' of the Milky
Way in hierarchical-clustering models (e.g. Lacey \& Cole 1993). Thus
one may use inferences from the mass distribution of the inner Galaxy
to constrain the fate of substructure back to that time.

The mean density interior to the periGalactic approach ($\sim 1$~kpc)
of the `low-alpha' stars is then $ \simgt 2.5$~M$_\odot$~pc$^{-3}$. To
be robust against tides, a satellite galaxy on an orbit with this
small a periGalactic distance would need an mean internal density of
around three times this, or $\simgt 7$~M$_\odot$~pc$^{-3}$. This value
is significantly larger (by orders of magnitude) than the typical mean
densities derived for dwarf galaxies, both gas-poor and gas-rich
(e.g. Puche \& Carignan 1991; Ibata et al.  1997).  Thus essentially
all dwarf galaxies as we see them now (presumably the survivors) would
be disrupted after only a couple of such pericenter passages (see
Johnston, Hernquist \& Bolte 1996).  Indeed, they would likely be
disrupted significantly earlier in their orbital decay, when the
pericenter distances are of order the present derived apocenter
distances of the anomalous halo stars (e.g. Ibata et al.  1997).

Thus, if the halo stars under discussion represent stars captured from
a satellite galaxy, that galaxy had to be sufficiently massive to
spiral into the center of the Milky Way in less than a Hubble time --
this implies a mass perhaps 10\% that of the older stars in the
central Milky Way, or equal to the entire mass of the stellar halo
(e.g. Carney, Latham \& Laird 1990).  This satellite had also to be
significantly more robust than any of the extant satellite galaxies
(the survivors).  If this satellite contributed to the formation of
the thick disk, it did so without donating any of its own stars to the
disk. Each of these is rather surprising.

\subsection{Halo Star-Formation Regions}

The requirements for the progenitor structures for the metal-rich,
`low-alpha' halo stars may be summarised as being dense enough to
survive to a periGalactic distance of $\sim 1$~kpc, and have a deep
enough local potential well to self-enrich to $-1$~dex, while
sustaining star formation for $\sim 1$~Gyr, being robust not only to
internal feedback from star formation, but also external effects.
Plausibly the first two of these requirements are satisfied by a
density criterion for the substructure, given that the escape energy
per unit mass scales as $\rho \, r^2$.  Sustaining star formation
requires gas retention, and the further requirement that the effects
of collisions with other star-forming clouds, or passages through the
(forming) gas disk of the Galaxy, be minimised.

As discussed by many authors (see e.g. Gunn 1980; Jones, Palmer \&
Wyse 1981) gravitational instability in the cooling layers behind
shocked gas in the haloes of protogalaxies produces systems of density
$\simgt 50$M$_\odot$pc$^{-3}$ and of characteristic (Jeans) mass of a
few times $10^6$M$_\odot$. A more sophisticated analysis by Fall \&
Rees (1985) included thermal instability in the hot, gaseous halo of a
proto-galaxy, like the Milky Way, to produce cool, dense regions, that
are seeds for gravitational instability.  Fall \& Rees (1985) argue
that a fairly massive proto-galaxy, with circular velocity greater
than $\sim 150$~km/s, is required for this scenario (essentially to
provide the hot phase of the gaseous halo). In flat Cold-Dark-Matter
dominated cosmologies (e.g. White \& Frenk 1991), dark haloes of this
circular velocity have peak abundance at redshifts $\simgt 3$, or
lookback times of $\simgt 12$~Gyr. Thus this picture of fragmentation
inside dark haloes is reasonable even in hierarchical clustering
scenarios.

The fragments undergo star formation and are subsequently dispersed,
either by the internal effects of star formation, or the external
effects of cloud-cloud collisions or tidal disruption by the Galactic
potential. It is plausible that the fragments probing smaller
Galactocentric distance have larger internal densities (indeed, as
predicted by the model of Fall \& Rees 1985) and larger escape
velocities, allowing greater self-enrichment and thus providing a
relation between R$_{peri}$ and [Fe/H] as observed in the remnant
stars, now the field halo (higher metallicity for lower R$_{peri}$;
Ryan \& Norris 1991).  The most dense regions could plausibly form
globular clusters, as suggested by Fall \& Rees (1985), even some of
which are subsequently destroyed (e.g. Gnedin \& Ostriker 1997).

The bias to large apoGalactic distances for the orbits of the
`low-alpha' stars follows naturally from the expected longer survival
times of large, self-enriching clouds in the outer regions of
proto-Galaxy, due to the longer collision times there, and the larger
orbital times.  Additionally, a large-scale density gradient could
provide slower local star formation rates in the outer regions,
provided the properties of density perturbations correlate with their
local background.

Thus fragmentation would appear to provide the kind of substructure
required by the observations.

\section { Disk -- Halo (Dis)connection}

A striking aspect of the data in Figure 1 is that the value of the
`Type~II plateau' of element ratios is the same for the halo stars as
for the disk stars of the same metallicity. Following the same logic
as Wyse \& Gilmore (1992), this implies that the IMF of the (massive)
stars that enriched these stars was the same in the early disk and
halo. Inversion of the value of the element ratios to determine the
actual slope of the massive star IMF requires adoption of theoretical
supernova yields and is hence uncertain, but the robust statement that
the IMF was invariant in the enrichment of disk and halo may be made
immediately.

It is also clear from Figure 1 that most of the metal-rich halo stars
do not have element ratios equal to this `Type II plateau', and that
the {\it typical\/} halo and disk stars of equal metallicity ([Fe/H] $
\sim -1$~dex) do {\it not\/} have the same values of the element
ratios.  There is not the smooth continuity in chemical enrichment
from halo to disk often derived in models of chemical evolution.  Of
course, these halo stars are on orbits of very different angular
momentum than those of the disk stars, further emphasizing the
distinction between local samples of disk and halo, and the low
likelihood that the halo, as sampled, pre-enriched the disk (see Wyse
\& Gilmore 1992; Gilmore 1995).  There is also a spread in [Fe/H] at
fixed values of [Mg/Fe]; as mentioned earlier in section 3, this may
be achieved by different gas flows/mixing and star formation
histories.

The onset of star formation in the local disk, as measured by the
white dwarf luminosity function, has most recently been estimated to
be 10--12~Gyr ago (an increase over earlier estimates, in part due to
the inclusion of phase separation and crystallisation energy in the
white dwarf cooling curves; Hernanz et al.  1994; Chabrier 1998).
This older age is in agreement with the ages of the oldest open
clusters (e.g. Phelps 1997) and the age of the thick disk (e.g.
Gilmore, Wyse \& Jones 1995) and in better agreement with the ages of
individual stars from Edvardsson et al.  (1993; see also Ng \&
Bertelli 1998 for post-Hipparcos age estimates for stars from this
sample).  The recent downward revision in the ages of the halo
globular clusters to $\sim 12$~Gyr, based on Hipparcos calibrations of
the luminosities of field subdwarfs (e.g.  Gratton et al.  1997b; Reid
1998), leaves little room for a significant hiatus between the epoch
of halo star formation and the onset of star formation in the disk
(however, this calibration depends on many corrections and is not
unique -- see e.g. Pont et al. 1998).  It would follow that disk and
halo were independent entities from the very early stages of Galaxy
formation.

\section {Conclusions}

 The recent observations of halo stars with elemental abundances
reflecting pre-enrichment from intermediate-mass stars and Type Ia
supernovae can be understood in many models of halo formation.  The
establishment of trends with kinematics, suggested at in the present
data, will be important in distinguishing models, as will be age
determinations. The fact that the `low-alpha' halo stars tend to be
metal-rich favors self-enrichment scenarios, and the survival of
individual star-forming complexes for approximately a Gyr.  That the
observed low-alpha stars are on orbits that plunge close to the
Galactic center also implies that a reasonably robust parent system is
required. We argue for an interpretation that simply appeals to a
small variation in the duration of star formation within regions of
the stellar halo, rather than `distinct accretion events'.

The present data implicates a significant fraction of the metal-rich,
outer halo.  One may normalise to the stellar halo as a whole, using
the metallicity distribution for local halo stars of Carney et al.
(1996) in which $\sim 15$\% of the stellar halo has [Fe/H]$ \simgt
-1$~dex, combined with the halo structural parameters of Unavane, Wyse
\& Gilmore (1996) in which $\sim 10$\% of the halo is located beyond
$Z = 5$~kpc, to conclude that at most a few percent (10\% times 15\%)
of the stellar halo is being discussed.  More data -- not restricted
to detailed elemental abundances for selected halo stars -- will
clearly help to substantiate these estimates.  For example, the
metallicity distribution of the halo is derived from local,
kinematically-selected samples, and the inner halo is poorly studied.
Further, there is currently little information about the {\it wings\/}
of the metallicity distribution of the stellar halo.

The present data for disk and halo stars of `overlapping'
metallicities [Fe/H]$ \sim -1$~dex show a disk--halo dichotomy, in
that the elemental abundances are different for typical disk stars and
for typical halo stars in this metallicity range, with the halo stars
having lower values of [$\alpha$/Fe] than the disk stars. This
complicates the requirements for chemical evolution models for the
disk that appeal to gas inflow from the halo; indeed if the halo had
ejected gas that pre-enriched the disk, one might have expected the
opposite situation, with the disk stars having lower values of element
ratios than the halo stars.  A simplification of chemical evolution
models is that the elemental abundance data favor a fixed massive-star
IMF for both disk and halo.

\acknowledgements{
We are grateful to NATO Scientific Affairs for a grant to aid our
collaboration.  RFGW thanks all at the Center for Particle Astrophysics (UC
Berkeley) for hospitality during the early stages of writing of this paper. 
She acknowledges partial support from NASA, through ATP grant NAG5-3928.}

\pagebreak

\end{document}